\documentclass[a4paper,11pt]{article}
\usepackage{jheppub}
\usepackage[latin1]{inputenc} 
\usepackage[T1]{fontenc}
\usepackage[safe]{textcomp}
\usepackage{lmodern} 
\usepackage{epsfig}
\usepackage{float}
\usepackage{amstext}  
\usepackage{amsfonts}
\usepackage{amsthm}
\usepackage{bm}       
\usepackage[bottom]{footmisc} 
\usepackage{array}                
\usepackage{xcolor} 
\usepackage{framed}
\usepackage{slashed}
\usepackage[absolute]{textpos}
\usepackage{axodraw4j}
\usepackage{dsfont}
\usepackage{multirow}

\newcommand{\p}{\partial}

\newcommand{\f}[2]{\frac{#1}{#2}}
\newcommand{\sss}[1]{\scriptscriptstyle#1}
\newcommand{\ssst}[1]{\scriptscriptstyle{\text{#1}}}

\newcommand{\bea}{\begin{eqnarray}}
\newcommand{\eea}{\end{eqnarray}}
\newcommand{\be}{\begin{equation}}
\newcommand{\ee}{\end{equation}}
\newcommand{\ba}{\begin{align}}
\newcommand{\ea}{\end{align}}
\newcommand{\beas}{\begin{eqnarray*}}
\newcommand{\eeas}{\end{eqnarray*}}
\newcommand{\bes}{\begin{equation*}}
\newcommand{\ees}{\end{equation*}}
\newcommand{\bas}{\begin{align*}}
\newcommand{\eas}{\end{align*}}

\newcommand{\eps}{{\varepsilon}}
\newcommand{\cd}{{\cdot}} 
\newcommand{\cf}{C_{\scriptscriptstyle{F}}} 
\newcommand{\ca}{C_{\scriptscriptstyle{A}}}
\newcommand{\tr}{T_{\scriptscriptstyle{F}}}

\newcommand{\dR}{d_{\scriptscriptstyle{R}}}
\newcommand{\Ng}{n_{\scriptscriptstyle{g}}}

\newcommand{\Nc}{N_{\scriptscriptstyle{c}}}
\newcommand{\Nf}{n_{\scriptscriptstyle{f}}}
\newcommand{\gs}{g_{\scriptscriptstyle{s}}}

\newcommand{\als}{\alpha_{\scriptscriptstyle{s}}}
\newcommand{\as}{a_{\scriptscriptstyle{s}}}
\newcommand{\lb}{\left(}
\newcommand{\rb}{\right)}

\definecolor{bluemar}{rgb}{0,0,.5}
\definecolor{redmar}{rgb}{.8,0,0}
\definecolor{greenmar}{rgb}{0,.5,0}

\def\bbuildrel#1_#2^#3%
{\mathrel{\mathop{\kern 0pt#1}\limits_{#2}^{#3}}}

\newcommand{\ice}[1]{\relax}
\newcommand{\beq}{\begin{equation}}
\newcommand{\eeq}{\end{equation}}

\allowdisplaybreaks[1]

\title{OPE of the energy-momentum tensor correlator and the gluon condensate operator in massless QCD to three-loop order}
\author[a]{M. F. Zoller}
\affiliation[a]{Institut f\"ur Theoretische Teilchenphysik, Karlsruhe
  Institute of Technology (KIT), \mbox{D-76128 Karlsruhe, Germany}}

\emailAdd{max.zoller@kit.edu}

\abstract{
The correlator of two gluonic operators plays an important role for example in transport properties of a 
Quark Gluon Plasma (QGP) or in sum rules for glueballs.

In \cite{Zoller:2012qv} an operator product expansion (OPE) at zero temperature was performed
for the correlators of two scalar operators \mbox{$O_1=-\f{1}{4} G^{\mu \nu}G_{\mu \nu}$}
and two QCD energy-momentum tensors $T^{\mu\nu}$. There we presented analytical two-loop results for the Wilson coefficient $C_1$ in front of the gluon 
condensate operator $O_1$. In this paper these results are extended to three-loop order.

The three-loop Wilson coefficient $C_0$ in front of the unity operator $O_0=\mathds{1}$ was already presented in 
\cite{Zoller:2012qv} for the \mbox{$T^{\mu\nu}$-correlator.}
For the \mbox{$O_1$-correlator} the coefficient $C_0$ is known to four loop order from \cite{Baikov:2006ch}. For the correlator of two pseudoscalar 
operators \mbox{$\tilde{O}_1=\eps_{\mu\nu\rho\sigma} G^{\mu \nu} G^{\rho \sigma}$} both coefficients $C_0$ and $C_1$ were computed in 
\cite{Zoller:2013ixa} to three-loop order. At zero temperature $C_0$ and $C_1$ are the leading Wilson coefficients in massless QCD.
}

\arxivnumber{}
\subheader{TTP14-023 \\SFB/CPP-14-56}

\begin{document}
\maketitle

\setlength{\fboxrule}{0.5 mm} 

\section{Introduction and definitions}

Correlators of two local operators $O(x)$ are important objects in quantum field theory. In momentum space they are defined as
\be 
i\int\!\mathrm{d}^4x\,e^{iqx} T\{\,[O](x)[O](0)\}{},
\label{corrOP}
\ee
where $[O]$ ist defined to be a renormalized version of the operator $O$, i.e.~matrix elements of $[O]$ are finite.\footnote{By a local operator $O$ we mean
a combination of fields at the same space-time point. The bare operator $O^{\ssst{B}}$ is the same combination but with bare fields and in the simplest case
$[O]=Z^{O} O^{\ssst{B}}$ is the renormalized operator with a renormalization constant $Z^{O}$. In some cases a set of operators mixes under renormalization giving $[O_i]=Z^{O}_{ij} O_j^{\ssst{B}}$,
where all $[O_i]$ are finite if inserted into a Greens function. If more than one operator is inserted into a Greens function additional divergences may
appear if these operators are taken to be at the same space-time point. 
Such contributions are called contact terms.
}
For sum rules we are usually interested in the vacuum expectation value (VEV) of the correlator
\be 
\Pi(Q^2)=i\int\!\mathrm{d}^4x\,e^{iqx}\,\langle 0|T\{\,[O](x)[O](0)]\}|0 \rangle \qquad
\label{corrVEV}
\ee
with large $Q^2:=-q^2>0$, i.e.~in the Euclidean region of momentum space.
The function $\Pi(Q^2)$ is connected to the spectral density $\textbf{Im}\Pi(s)$ in the region of 
physical momenta through a dispersion relation (see e.g.~\cite{forkel_sumrule}).

The leading contribution to $\Pi(Q^2)$ can be computed perturbatively and is exactly the first Wilson coefficient in front of the unity operator $O_0=\mathds{1}$.
In order to include non-perturbative effects as well the correlator \eqref{corrOP} is expanded in a series of local operators with Wilson coefficients containing the dependence
on $q$ in momentum space or $x$ in x-space \cite{wilson_ope}. This operator product expansion (OPE) 
has the form
\bea 
i\int\!\mathrm{d}^4x\,e^{iqx} T\{\,[O](x)[O](0)\}
&=&\sum \limits_i  (Q^2)^\f{2 \text{ dim}(O)-\text{dim}(O_{i})-4}{2} C_{i}^{\ssst{B}}(q) O_{i}^{\ssst{B}}
\label{OPE_B}\\
&=&\sum \limits_i  (Q^2)^\f{2 \text{ dim}(O)-\text{dim}(O_{i})-4}{2} C_{i}(q) [O_{i}]{},
\label{OPE_R}
\eea
where the index B marks bare quantities. The factor $(Q^2)^\f{2 \text{ dim}(O)-\text{dim}(O_{i})-4}{2}$ is 
constructed from the mass dimensions of the operators in order to make $C_i(q)$ dimensionless.

The perturbative contribution is separated from the non-perturbative condensates 
in an operator product expansion (OPE) and hence resides in the Wilson coefficients in front of local operators.
These Wilson coefficients are calculated perturbatively using the method of projectors \cite{Gorishnii:1983su,Gorishnii:1986gn}
and contain the perturbative contribution to the correlator in question. 
If we insert expansion \eqref{OPE_R} into \eqref{corrVEV} we are left with the task of determining 
the VEVs of the local operators $[O_{i}]$, the so-called condensates
\cite{Shifman:1978bx}, which contain the non-perturbative part.
These need to be derived from low energy theorems or be calculated on the lattice.

Three gluonic operators with the quantum numbers $J^{PC}=0^{++},0^{-+}$ and $2^{++}$ are usually considered:\footnote{For details 
on the sum rule approach to glueballs with the same quantum numbers see e.g.~\cite{forkel_sumrule}.
An OPE at one-loop level has been performed for the scalar \cite{Novikov_scalargluonium} and pseudoscalar \cite{Novikov:1979ux} correlator.
Recent discussions on glueballs using an OPE of these correlators can be found in \cite{Bochicchio:2013tfa,Bochicchio:2013aha}.
}
 \begin{align}
     O_1^{\ssst{B}}(x)  &=-\f{1}{4} G^{{\ssst{B}}\,a\,\mu \nu}G^{\ssst{B}\,a}_{\mu \nu}(x)  & \text{(scalar)} {},\label{O1}\\
     \tilde{O}_1^{\ssst{B}}(x)  &=\eps_{\mu\nu\rho\sigma}G^{{\ssst{B}}\,a\,\mu \nu}G^{^{\ssst{B}}\,a\,\rho \sigma}(x)  & \text{(pseudoscalar)}{},\label{O1t}\\ 
     O_T^{\mu \nu}(x) &=T^{\mu \nu}(x) \label{OT} & \text{(tensor)}{}
    \end{align} 
    with the bare gluon field strength tensor
\be
G_{\mu \nu}^{a\,\ssst{B}}=
\p_\mu A_\nu^{a\,\ssst{B}} - \p_\nu A_\mu^{a\,\ssst{B}} 
+ \gs^{\ssst{B}} f^{abc} A_\mu^{\ssst{B}\,b} A_\nu^{\ssst{B}\,c}
{},
\ee
where $f^{abc}$ are the structure constants and $T^a$ the generators of the SU($\Nc$) gauge group.
As described in \cite{Zoller:2012qv} for $T^{\mu \nu}$ we use  
the gauge invariant and symmetric energy-momentum tensor of (massless) QCD: 
\be
\begin{split}
T_{\mu \nu}|_{\sss{\mathrm{ginv}}}=
&-G^{\ssst{B}\,a}_{\mu \rho}G_{\nu}^{{\ssst{B}\,a}\,\rho}
+\f{i}{4}\bar{\psi}^{\ssst{B}}\left(\overleftrightarrow{\p_\mu} \gamma_\nu
+\overleftrightarrow{\p_\nu} \gamma_\mu\right)\psi^{\ssst{B}}
+\f{1}{2}\gs^{\ssst{B}}\bar{\psi}^{\ssst{B}}\left(A_\mu^{\ssst{B}\,a} T^a \gamma_\nu
+A_\nu^{\ssst{B}\,a}\ T^a \gamma_\mu\right)\psi^{\ssst{B}} \\
&-g_{\mu\nu}\left\{ -\f{1}{4}\,G_{\rho \sigma}^{\ssst{B}\,a} G^{{\ssst{B}\,a}\,\rho \sigma}
+\f{i}{2}\bar{\psi}^{\ssst{B}}\overleftrightarrow{\slashed{\p}}\psi^{\ssst{B}} 
+ \gs \bar{\psi}^{\ssst{B}}\slashed{A}^{\ssst{B}\,a} T^a\psi^{\ssst{B}} \right\}.
\end{split}
\label{5Tqcdginv}
\ee
In \cite{nielsen_Tmunu} it was argued that if we are only interested in matrix elements of only gauge invariant operators
it is not necessary to consider the ghost terms appearing in the full energy-momentum tensor of QCD. It was also proven that
the energy-momentum tensor of QCD is a finite operator without further renormalization.

The operator $O_1$ and the Wilson coefficients $C_1$, however, have to be renormalized in the following way:
\bea
[O_1]&=&Z_G O_1^{\ssst{B}}=-\f{Z_G}{4}G^{{\ssst{B}}\,a\,\mu \nu}G^{\ssst{B}\,a}_{\mu \nu}{} \label{O1ren}\\
C_1&=&\f{1}{Z_G}C_1^{\ssst{B}}. \label{C1ren}
\eea
The renormalization constant
\be Z_G=1+\als\f{\p}{\p\als}\ln Z_{\als}=\lb 1- \f{\beta(\als)}{\eps}\rb^{-1} \label{ZGdef}\ee
was derived in  \cite{Nielsen:1975ph,Spiridonov:1984br} from the
renormalization constant $Z_{\als}$ for $\als$. At first order in $\als$ we find $Z_G=Z_{\als}$,
which is not true in higher orders however.
We take the definition
\beq\beta(\als)=
\mu^2\f{\mathrm{d}}{\mathrm{d}\mu^2}\, \ln \als
=  - \sum_{i \ge 0} \beta_i \, \left( \frac{\als}{\pi} \right)^{i+1}
{}
\label{be:def}
\eeq
for the $\beta$-function of QCD, which is available at four-loop level \cite{vanRitbergen:1997va,Czakon:2004bu}.
For the renormalization of $\tilde{O}_1^{\ssst{B}}$, which mixes with a pseudoscalar
fermionic operator under renormalization, and its OPE we refer to \cite{Larin:1993tq,Zoller:2013ixa}.

The correlators of $O_1$ and $O_T^{\mu \nu}$ have been discussed in \cite{Zoller:2012qv}, 
where $C_1$ has been presented at two-loop level.
The results of this work are derived within the same theoretical and methodical framework, 
which is why we can refer to this work for most technical details. 
$C_0$ is also known to three-loop level for the $T^{\mu \nu}$-correlator \cite{Zoller:2012qv} 
and at two-, three- and four-loop level for the $O_1$-correlator from
\cite{Kataev:1982gr},\cite{Chetyrkin:1997iv} and \cite{Baikov:2006ch} respectively.
Three-loop results for $C_0$ and $C_1$ for the correlator of two operators $\tilde{O}_1$ have been derived in \cite{Zoller:2013ixa}.

The VEV of the energy-momentum tensor correlator 
\bea 
T^{\mu\nu;\rho\sigma}(q)&:=&\langle 0|\hat{T}^{\mu\nu;\rho\sigma}(q)|0\rangle \label{TTvev},\\
\hat{T}^{\mu\nu;\rho\sigma}(q) &:=& i\int\!\mathrm{d}^4x\,e^{iqx}\,T\left\{ T^{\mu\nu}(x)T^{\rho\sigma}(0)\right\} \label{TThat}
\eea
is an important quantitiy in calculations of transport properties of a Quark Gluon Plasma (QGP), such as
the shear viscosity of the plasma (see e.g. \cite{Meyer:2008sn,Meyer:2007ic}) and spectral functions for some tensor channels in the QGP \cite{Meyer:2008gt}.

The correlator \eqref{TThat} is linked to the $O_1$-correlator
\be
Q^4\, \Pi^{\ssst{GG}}(q^2) := i\int \! \mathrm{d}^4x\,e^{iqx}\,
\langle 0|T\,[O_1 (x) O_1(0)]|0\rangle
\label{O1O1:def}
\ee
through the trace anomaly \cite{Collins:1976yq,nielsen_Tmunu}
\be \label{Tmumu_G2}
T^\mu_{\,\,\mu}=\f{\beta(a_s)}{2} \, [G^a_{\rho\sigma} G^{a\,\rho\sigma}]
= -2\,\beta(a_s) \, [O_1]
{},
\ee
which leads to
\beq
g_{\mu\nu} g_{\rho\sigma} T^{\mu\nu;\rho\sigma}(q)= 4\, \beta^2(\als) \,Q^4\, \Pi^{\ssst{GG}}(q^2) + \mbox{contact terms}.
\label{expect}
\eeq

Both correlators $\Pi^{\ssst{GG}}$ and $T^{\mu\nu;\rho\sigma}(q)$ have been studied 
in hot Yang-Mills theory in many works, see e.g.~\cite{Laine:2013vpa,Kajantie:2013gab,Zhu:2012be,Laine:2010tc,Schroder:2011ht} and references therein.

At zero temperature \eqref{TThat} has the asymptotic behaviour
\be \hat{T}^{\mu\nu;\rho\sigma}(q)
\bbuildrel{=\!=\!=}_{q^2 \to -\infty}^{} 
C_{0}^{\mu\nu;\rho\sigma}(q) \mathds{1} + C_{1}^{\mu\nu;\rho\sigma}(q) [O_1]+\ldots 
\label{TTexp1}\ee
where the tensor structure of the correlator resides in the Wilson coefficients if we are ultimately only interested
in the VEV of the correlator.

Local tensor operators can always be decomposed in a trace part and a traceless part, i.e.~for two Lorentz indices
\be O^{\mu\nu}=\underbrace{O^{\mu\nu}-\f{1}{D}g^{\mu\nu} O^{\rho}_{\,\,\rho}}_{\text{traceless part}}
+\underbrace{\f{1}{D}g^{\mu\nu} O^{\rho}_{\,\,\rho}}_{\text{trace part}} \ee
where $D$ is the dimension of the space time.
The VEV of the traceless part vanishes due to the Lorentz invariance of the vacuum and
only a local scalar operator $O^{\rho}_{\,\,\rho}$ survives.

The OPE of the correlator \eqref{O1O1:def} reads
\beq
 Q^4\, \Pi^{\ssst{GG}}(q^2) \bbuildrel{=\!=\!=}_{q^2 \to -\infty}^{}   C_0^{\ssst{GG}} \,Q^4 
\ \ + \ \  C_1^{\ssst{GG}} \, \langle 0 |[O_1]| 0\rangle.
\eeq

\section{Calculation and results}

As discussed in \cite{Zoller:2012qv} there are five independent tensor structures for \eqref{TTexp1} allowed by the symmetries
$\mu \longleftrightarrow \nu$, $\rho \longleftrightarrow \sigma$ and $(\mu \nu) \longleftrightarrow (\rho \sigma)$ of \eqref{TTexp1}.
These are
\be \begin{split}
t_1^{\mu\nu;\rho\sigma}(q)&=q^\mu q^\nu q^\rho q^\sigma ,\\
t_2^{\mu\nu;\rho\sigma}(q)&=q^2 \ \lb q^{\mu} q^{\nu} g^{\rho \sigma}+q^{\rho} q^{\sigma} g^{\mu \nu} \rb ,\\
t_3^{\mu\nu;\rho\sigma}(q)&=q^2 \ \lb q^{\mu} q^{\rho} g^{\nu \sigma}+q^{\mu} q^{\sigma} g^{\nu \rho} 
  +q^{\nu} q^{\rho} g^{\mu \sigma}+q^{\nu} q^{\sigma} g^{\mu \rho} \rb ,\\
t_4^{\mu\nu;\rho\sigma}(q)&=\lb q^2 \rb^2g^{\mu\nu}g^{\rho\sigma} ,\\
t_5^{\mu\nu;\rho\sigma}(q)&=\lb q^2 \rb^2 \lb g^{\mu \rho} g^{\nu \sigma}+g^{\mu \sigma} g^{\nu \rho}\rb.     
    \end{split} 
\label{t1_t5}
\ee
Due to the fact that the energy-momentum tensor is conserved except for contact terms, i.e.
\be 
q_\mu\,T^{\mu\nu;\rho\sigma}(q)= \text{local contact terms},
\label{Tmumu:conservation}
\ee
and due to the irrelevance of these contact terms for physical applications we can reduce
\eqref{t1_t5} to only two independent tensor structures, which have already been suggested in \cite{Pivovarov_tensorcurrents},
after contact term subtraction:
:
\be \begin{split}
                        t_S^{\mu\nu;\rho\sigma}(q)=&\eta^{\mu\nu} \eta^{\rho\sigma} \\
t_T^{\mu\nu;\rho\sigma}(q)=&\eta^{\mu\rho} \eta^{\nu\sigma}
                      +\eta^{\mu\sigma} \eta^{\nu\rho}
                      -\f{2}{D-1}\eta^{\mu\nu} \eta^{\rho\sigma} \\[2ex]
\text{with} \quad 
\eta^{\mu\nu}(q)=&q^2 g^{\mu\nu} - q^\mu q^\nu{}.
\label{tST}
                       \end{split} \ee

The structure $t_T^{\mu\nu;\rho\sigma}(q)$ is traceless and orthogonal to $t_S^{\mu\nu;\rho\sigma}(q)$. 
Hence the latter corresponds to the part coming from the traces of the energy-momentum tensors.
The Wilson coefficient in front of the local operator $[O_1]$ has the form
\be \begin{split}
C_{1}^{\mu\nu;\rho\sigma}(q)=&
\sum \limits_{r=1,5} \,t_r^{\mu\nu;\rho\sigma}(q)\,\f{1}{(Q^2)^{2}}\,C_{i}^{(r)}(Q^2)\\
=&
\sum \limits_{r=T,S} \,t_r^{\mu\nu;\rho\sigma}(q)\,\f{1}{(Q^2)^{2}}\,C_{i}^{(r)}(Q^2) \;\;(+\text{ contact terms)}.  
    \end{split} \label{Ci2ten} \ee
where the contact terms have to be $\propto t_4^{\mu\nu;\rho\sigma}(q)$ or $\propto t_5^{\mu\nu;\rho\sigma}(q)$
as $t_r^{\mu\nu;\rho\sigma}(q)\,\f{1}{(Q^2)^{2}}$ is not local for $r \in\{1,2,3\}$. This was checked explicitly in our three-loop
result.

Just like in \cite{Zoller:2012qv} (see this paper for more details) 
the method of projectors \cite{Gorishnii:1983su,Gorishnii:1986gn} was used
in order to compute the coefficient $C_1^{\mu\nu;\rho\sigma}(q)$. 
We apply the same projector to both sides of \eqref{OPE_B}:
\be 
{\bf P}\{i\int\!\mathrm{d}^4x\,e^{iqx} T\{\,[O](x)[O](0)\}\}=
\sum \limits_i C_{i}^{{\ssst{B}}}(q)\, {\bf P}\{O_{i}^{\ssst{B}}\} \stackrel{!}{=} C_{1}^{\ssst{B}}(q).\\
 \label{proj2OO}
\ee
The projector ${\bf P}$ is constructed in such a way that it 
maps every operator on the rhs of \eqref{OPE_B} to zero except for $O_{1}^{\ssst{B}}$, which is mapped to $1$ and hence gives
us the bare Wilson coefficient $C_{1}^{\ssst{B}}$ on the lhs.
For the $T^{\mu\nu}$-correlator \eqref{TThat} this is done
after contracting the free Lorentz indices with a tensor $\tilde{t}^{(r)}_{\mu\nu;\rho\sigma}(q)$ composed of the momentum  
$q$ and the metric $g^{\mu\nu}$ in order to get the scalar pieces in \eqref{Ci2ten}:\footnote{
The $\tilde{t}^{(r)}_{\mu\nu;\rho\sigma}$ can be constructed as linear combinations of the
$t_r^{\mu\nu;\rho\sigma}(q)$ in \eqref{t1_t5}.}
\be 
{\bf P}\{\tilde{t}^{(r)}_{\mu\nu;\rho\sigma}(q)T^{\mu\nu;\rho\sigma}(q)\}
=\sum \limits_i C_{i}^{{\ssst{B}},(r)}(Q^2)\, {\bf P}\{O_{i}^{\ssst{B}}\}.\\
 \label{TTexp3}
\ee
We use the following projector:\footnote{The Feynman diagram has been drawn with the 
Latex package Axodraw \cite{Vermaseren:1994je}.}
\be
C_{1}^{\ssst{B}}(q) =\f{\delta^{ab}}{\Ng}\f{g^{\mu_1 \mu_2}}{(D-1)}
\f{1}{D}\f{\p}{\p k_1} \cd \f{\p}{\p k_2} \left. \left[
  \begin{picture}(165,50) (0,0)
    \SetWidth{0.5}
    \SetColor{Black}
    \Gluon(10,0)(50,0){5.5}{4.5}
    \Gluon(110,0)(150,0){5.5}{4.5}
    \LongArrow(35,15)(25,15)
    \LongArrow(125,15)(135,15)
\DashCArc(80,0)(56,45,135){4}
\Photon(80,0)(120,40){3}{4}
\Photon(80,0)(40,40){3}{4}
    \CCirc(80,0){30}{Black}{Blue}
    \SetColor{Red}
\Vertex(110,0){4}
\Vertex(50,0){4}
    \SetColor{Black}
    \Text(25,17)[lb]{\Large{\Black{$k_1$}}}
    \Text(125,17)[lb]{\Large{\Black{$k_2$}}}
    \Text(36,-17)[lb]{\Large{\Red{$g_B$}}}
    \Text(110,-17)[lb]{\Large{\Red{$g_B$}}}
    \Text(150,-20)[lb]{\Large{\Black{$\mu_2$}}}
    \Text(5,-20)[lb]{\Large{\Black{$\mu_1$}}}
    \Text(5,10)[lb]{\Large{\Black{$a$}}}
    \Text(150,10)[lb]{\Large{\Black{$b$}}}
  \end{picture}
\right] \right|_{k_i=0}
{}, \label{projectorC1pic}
\ee
where the blue circle represents the sum of all bare Feynman diagrams
which become 1PI after formal gluing (depicted as a dotted line in \eqref{projectorC1pic}) of
the two external lines representing the operators on the lhs of the OPE. These external legs
carry the large Euclidean momentum q. 

In order to produce all possible Feynman diagrams we
have used the program QGRAF \cite{QGRAF}.  These 
propagator-like diagrams were computed with the FORM \cite{Vermaseren:2000nd,Tentyukov:2007mu}
package MINCER \cite{MINCER} after projecting them to scalar
pieces. For the colour factors of the diagrams the FORM package COLOR
\cite{COLOR} was used.

We now give the three-loop results for the Wilson coefficient $C_1$ of the correlators \eqref{TThat} and \eqref{O1O1:def}
in the $\overline{\text{MS}}$-scheme. In the following the abbreviations
\mbox{$\as=\f{\als}{\pi}=\f{\gs^2}{4\pi^2}$} and \mbox{$l_{\sss{\mu q}}=\ln\lb\f{\mu^2}{Q^2}\rb$} are used,
where $\mu$ is the $\overline{\text{MS}}$ renormalization scale.
The number of active quark flavours is denoted by $\Nf$.
Furthermore, $\cf$ and $\ca$ are the quadratic Casimir
operators of the quark and the adjoint representation of the gauge group,
$\dR$ is the dimension of the quark representation, $\Ng$ is the number of gluons (dimension of the adjoint representation),
$\tr$ is defined through the relation \mbox{$\textbf{Tr}\lb T^a T^b\rb=\tr \delta^{ab}$}  
for the trace of two group generators.\footnote{For an SU$(N)$ gauge group these are $\dR=N$,
$\ca=2\tr N$ and $\cf=\tr\lb N-\f{1}{N}\rb$.\\ For  QCD (SU$(3)$) this means $\cf =4/3\,,\, \ca=3\,,\,\tr=1/2$ and $\dR = 3$.}

\bea
   C_{1}^{(S)} =&{}& \as \left\{\frac{22 \ca}{27}-\frac{8 \Nf \tr}{27}\right\} \nonumber\\
    &+& \as^2 \left\{\frac{83 \ca^2}{324}-\frac{8 \ca \Nf \tr}{81}-\frac{2 \cf \Nf \tr}{9}-\frac{4 \Nf^2   \tr^2}{81}\right\}
    \nonumber\\
   &+& \as^3 \left\{-\frac{466 \ca^3}{729}
   +\frac{1309 \ca^2 \Nf \tr}{1944}
   -\frac{7}{648} \ca \cf \Nf \tr
   -\frac{313}{972} \ca \Nf^2 \tr^2 \right.\label{C1Sfin}\\&{}&\left. 
   +\frac{1}{36}   \cf^2 \Nf \tr  
   -\frac{7}{162} \cf \Nf^2 \tr^2
   +\frac{20 \Nf^3 \tr^3}{729}\right.\nonumber\\&{}&\left. 
      +l_{\sss{\mu q}} \left(
   -\frac{1331   \ca^3}{3888}
   +\frac{121}{324} \ca^2 \Nf \tr
   -\frac{11}{81} \ca \Nf^2 \tr^2
   +\frac{4 \Nf^3   \tr^3}{243}\right)
   \right\}\nonumber 
\eea

\bea
   C_{1}^{(T)} =&{}&  \as \left\{-\frac{5 \ca}{18}-\frac{5 \Nf \tr}{72}\right\}\nonumber\\
   &+&\as^2 \left\{-\frac{83   \ca^2}{432}+\frac{41 \ca \Nf \tr}{432}
   +\frac{43 \cf \Nf \tr}{96}-\frac{\Nf^2   \tr^2}{216}\right\}\nonumber\\
   &+&\as^3 \left\{
   -\frac{3 \ca^3 \zeta_3}{8}
   +\frac{103 \ca^3}{15552}
   -\frac{27}{80} \ca^2 \Nf \tr   \zeta_3
   +\frac{72239 \ca^2 \Nf \tr}{103680}  \right.\label{C1Tfin}\\&{}&\left.  
   +\frac{3}{8} \ca \cf \Nf \tr   \zeta_3
   +\frac{923 \ca \cf \Nf \tr}{1728}
   -\frac{3}{40} \ca \Nf^2 \tr^2 \zeta_3
   -\frac{217   \ca \Nf^2 \tr^2}{1620} \right.\nonumber\\&{}&\left.
   -\frac{241}{768} \cf^2 \Nf \tr   
   -\frac{21}{40} \cf \Nf^2 \tr^2   \zeta_3
   +\frac{929 \cf \Nf^2 \tr^2}{17280}
   +\frac{5 \Nf^3 \tr^3}{1944}  \right.\nonumber\\&{}&\left.
      +l_{\sss{\mu q}} \left(\frac{107 \ca^3}{5184}
   +\frac{73}{864} \ca^2   \Nf \tr
   +\frac{131}{384} \ca \cf \Nf \tr
   -\frac{7}{108} \ca \Nf^2 \tr^2 \right.\right.\nonumber\\&{}&\left.\left.
   -\frac{1}{6}   \cf \Nf^2 \tr^2
   +\frac{\Nf^3 \tr^3}{648}\right)
   \right\}\nonumber 
\eea

In \cite{Zoller:2012qv} it was shown that up to two-loop level
the coefficient $C_{1}^{(S)}$, which corresponds to the trace of the two energy-momentum tensors
in the correlator \eqref{TThat}, can be written in the form
\beq
C_{1}^{(S)}
= -\frac{8}{9}\,\beta(\as)\, \left(1+ \f{\beta(a_s)}{2} \right) + {\cal O}(\als^3)
{},
\eeq
where the first factor $\beta(\as)$ is due to the trace anomaly \eqref{Tmumu_G2}.
It is interesting to check whether we can find a similar structure in terms of the $\beta$-function
at three-loop level.
However, we do not find such an elegant representation at the next loop order. The closest we get is
\beq \begin{split}
C_{1}^{(S)}
=& -\frac{8}{9}\,\beta(\as)\, \left[1+ \f{\beta(a_s)}{2} - 
\left(\f{5}{3}+l_{\sss{\mu q}}\right) \f{\beta(a_s)^2}{2} \right. \\ &\left.+
\as^2 \left(-\frac{5 \ca^2}{18}
+\frac{13 \ca \Nf \tr}{72}
+\frac{\cf \Nf   \tr}{8} \right)\right]
+ {\cal O}(\als^4)
{}. \end{split}
\eeq

>From the renormalization group invariance (RGI) of the energy-momentum tensor and \eqref{Tmumu_G2} follows
that RGI invariant Wilson coefficients for RGI operators on the rhs of the OPE \eqref{OPE_R}
can be constructed as already explained in \cite{Zoller:2012qv}.
The scale invariant version of the operator $O_1$ is defined by 
\beq
O_1^{\ssst{RGI}} := \hat{\beta}(a_s) \, [O_1], \ \ \  \hat{\beta}(a_s) = \frac{-\beta(a_s)}{\beta_0} = 
a_s\left(1+  \sum_{i \ge 1} \frac{\beta_i}{\beta_0} a_s^i \right)
{}. \label{O1RGI}
\eeq
>From this and the scale invariance of the correlator \eqref{TThat} RGI Wilson coefficients can be defined as
\beq \begin{split}
C^{(S)}_{1,\ssst{RGI}} &:= C^{(S)}_{1}/\hat{\beta}(a_s),\\
C^{(T)}_{1,\ssst{RGI}} &:= C^{(T)}_{1}/\hat{\beta}(a_s),
\label{C1STRGI}
\end{split} \eeq
such that
\be C^{(S,T)}_{1,\ssst{RGI}} O_1^{\ssst{RGI}}=C^{(S,T)}_{1}[O_1]. \ee
We find
\bea
   C_{1,\ssst{RGI}}^{(S)} =&{}&
\frac{22 \ca}{27}-\frac{8 \Nf \tr}{27} \nonumber\\
&+&\as \left\{-\frac{121 \ca^2}{324}+\frac{22   \ca \Nf \tr}{81}-\frac{4 \Nf^2 \tr^2}{81}\right\}\nonumber\\
&+&\as^2 \left\{
-\frac{12661 \ca^3}{11664}
+\frac{365}{324} \ca^2 \Nf \tr
+\frac{11}{54} \ca \cf \Nf \tr \right.\label{C1SRGI_fin}\\&{}&\left.
-\frac{83}{243} \ca \Nf^2 \tr^2
-\frac{2}{27}   \cf \Nf^2 \tr^2 
+\frac{20 \Nf^3 \tr^3}{729} \right.\nonumber\\&{}&\left.
+l_{\sss{\mu q}} \left(
-\frac{1331   \ca^3}{3888}
+\frac{121}{324} \ca^2 \Nf \tr
-\frac{11}{81} \ca \Nf^2 \tr^2
+\frac{4 \Nf^3   \tr^3}{243}\right)
\right\}\nonumber
\eea
and
\bea
   C_{1,\ssst{RGI}}^{(T)} =&{}& 
   -\frac{5 \ca}{18}
   -\frac{5   \Nf \tr}{72}\nonumber\\
   &+&\as \left\{\f{1}{(11 \ca-4 \Nf \tr)}\left[
   \frac{107 \ca^3}{432 }
   +\frac{73 \ca^2 \Nf   \tr}{72 }
   -2 \cf \Nf^2 \tr^2                \right.\right.\nonumber\\&{}&\left.\left.
   +\frac{131   \ca \cf \Nf \tr}{32 }
   +\frac{\Nf^3 \tr^3}{54 }
   -\frac{7 \ca \Nf^2 \tr^2}{9 }\right]\right\}\nonumber\\
   &+& \as^2 \left\{
   -\frac{3}{40} \ca   \Nf^2 \tr^2 \zeta_3
   -\frac{217 \ca \Nf^2 \tr^2}{1620}             
   -\frac{241}{768} \cf^2 \Nf   \tr
   -\frac{21}{40} \cf \Nf^2 \tr^2 \zeta_3    \right.\nonumber\\&{}&\left.
   +\frac{929 \cf \Nf^2 \tr^2}{17280} 
   +\frac{5 \Nf^3   \tr^3}{1944}  
   -\frac{3 \ca^3   \zeta_3}{8}
   +\frac{103 \ca^3}{15552}    
   -\frac{27}{80} \ca^2   \Nf \tr \zeta_3 \right.\nonumber\\&{}&\left.
   +\frac{72239 \ca^2 \Nf \tr}{103680} 
   +\frac{3}{8} \ca \cf \Nf \tr   \zeta_3
   +\frac{923 \ca \cf \Nf \tr}{1728} \right.\label{C1TRGI_fin}\\&{}&\left.
   +\f{1}{(11 \ca-4 \Nf \tr)}\left[
   +\frac{1411 \ca^4}{864 }
   -\frac{509 \ca^3 \Nf \tr}{288 }
   -\frac{2525 \ca^2   \cf \Nf \tr}{576 } \right.\right.\nonumber\\&{}&\left.\left.
   +\frac{37 \ca^2 \Nf^2 \tr^2}{72 }
   +\frac{43 \cf^2 \Nf^2 \tr^2}{32 }
   -\frac{\cf \Nf^3 \tr^3}{72 }
   +\frac{727 \ca   \cf \Nf^2 \tr^2}{288 }  
   -\frac{5 \ca \Nf^3 \tr^3}{216 }   
   \right]\right.\nonumber\\&{}&\left.
   +\f{1}{(11 \ca-4 \Nf \tr)^2}\left[
   +\frac{53095 \ca^5}{5184 } 
   -\frac{308465 \ca^4 \Nf \tr}{20736 }
   +\frac{965 \ca^3 \cf   \Nf \tr}{864 }   \right.\right.\nonumber\\&{}&\left.\left. 
   +\frac{10255 \ca^3 \Nf^2 \tr^2}{3456 }
   +\frac{55 \ca^2 \cf^2 \Nf \tr}{48 }
   -\frac{65 \ca^2 \cf \Nf^2 \tr^2}{128 }
   +\frac{3175 \ca^2 \Nf^3 \tr^3}{5184 }   \right.\right.\nonumber\\&{}&\left.\left.
   -\frac{35   \cf^2 \Nf^3 \tr^3}{48 }
   -\frac{505 \ca \cf^2 \Nf^2 \tr^2}{192   }
   -\frac{55   \cf \Nf^4 \tr^4}{216 }
   -\frac{175 \ca \cf \Nf^3 \tr^3}{144 }  
   -\frac{395 \ca \Nf^4 \tr^4}{1296 }
   \right]\right.\nonumber\\&{}&\left.
   +l_{\sss{\mu q}} \left(\frac{107   \ca^3}{5184}   
   +\frac{73}{864} \ca^2 \Nf \tr
   +\frac{131}{384} \ca \cf \Nf \tr
   -\frac{7}{108}   \ca \Nf^2 \tr^2\right.\right.\nonumber\\&{}&\left.\left.
   -\frac{1}{6} \cf \Nf^2 \tr^2 
   +\frac{\Nf^3 \tr^3}{648}\right)              
   \right\}{}.\nonumber
\eea

In \cite{Zoller:2012qv} the three-loop logarithmic terms of \eqref{C1SRGI_fin} und \eqref{C1TRGI_fin} were constructed
from the two-loop result and the requirement that \mbox{$\mu^2\f{d}{d\mu^2} C^{(S,T)}_{1,\ssst{RGI}}$} vanishes identically.
and indeed we find the same result in this explicit calculation. This requirement also explains the absence of
Logarithms in the lower-order terms \cite{Zoller:2012qv}.
\bea
   C_{1}^{\ssst{GG}} = 
&-&1
+\as \left\{
-\frac{49 \ca}{36}+\frac{5 \Nf \tr}{9}
+l_{\sss{\mu q}} \left(\frac{\Nf \tr}{3}-\frac{11   \ca}{12}\right)
\right\}\nonumber\\
&+&\as^2 \left\{
   \frac{33 \ca^2 \zeta_3}{8}
   -\frac{11509 \ca^2}{1296}
   +\frac{3}{2} \ca \Nf \tr \zeta_3
   +\frac{3095 \ca \Nf   \tr}{648}
   -3 \cf \Nf \tr \zeta_3 \right.\nonumber\\&{}&\left.
   +\frac{13 \cf \Nf   \tr}{4}
   -\frac{25 \Nf^2 \tr^2}{81}
   +l_{\sss{\mu q}}   \left(
-\frac{1151 \ca^2}{216}
+\frac{97 \ca \Nf \tr}{27}      
+\cf \Nf \tr              \right.\right.\nonumber\\&{}&\left.\left.
-\frac{10 \Nf^2   \tr^2}{27}
\right)
   +l_{\sss{\mu q}}^2 \left(
   -\frac{121 \ca^2}{144}
   +\frac{11 \ca \Nf   \tr}{18}
   -\frac{\Nf^2 \tr^2}{9}
   \right)\right.\nonumber\\&{}&\left.
   +\f{1}{\eps}\left[
      -\frac{17 \ca^2}{24 }
      +\frac{5 \ca   \Nf \tr}{12 }
      +\frac{\cf \Nf \tr}{4 }\right]
   \right\}  \label{C1GG_3l_full}\\ 
&+&\as^3 \left\{
+\frac{5315 \ca^3   \zeta_3}{144}
-\frac{55 \ca^3 \zeta_5}{8}
-\frac{9775633 \ca^3}{186624}
-\frac{263}{144} \ca^2 \Nf \tr \zeta_3      \right.\nonumber\\&{}&\left.
-5   \ca^2 \Nf \tr \zeta_5                 
+\frac{1299295 \ca^2 \Nf \tr}{31104}
-\frac{331}{16} \ca \cf \Nf \tr \zeta_3
-\frac{15}{2} \ca   \cf \Nf \tr \zeta_5     \right.\nonumber\\&{}&\left.
+\frac{35707 \ca \cf \Nf \tr}{1152}            
-\frac{121}{36} \ca \Nf^2 \tr^2   \zeta_3 
-\frac{116773 \ca \Nf^2 \tr^2}{15552}
-9 \cf^2 \Nf   \tr \zeta_3     \right.\nonumber\\&{}&\left.
+15 \cf^2 \Nf \tr \zeta_5
-\frac{45}{16} \cf^2 \Nf \tr                
+\frac{13}{2} \cf \Nf^2 \tr^2   \zeta_3 
-\frac{2399}{288} \cf \Nf^2 \tr^2
+\frac{125 \Nf^3 \tr^3}{729}                  \right.\nonumber\\&{}&\left.
+l_{\sss{\mu q}} \left(\frac{363 \ca^3   \zeta_3}{32}
-\frac{360325 \ca^3}{10368}
+\frac{55757 \ca^2 \Nf \tr}{1728}
-\frac{33}{4} \ca \cf \Nf   \tr \zeta_3              \right.\right.\nonumber\\&{}&\left.\left.
+\frac{2527}{192} \ca \cf \Nf \tr 
-\frac{3}{2} \ca \Nf^2 \tr^2   \zeta_3
-\frac{2057}{288} \ca \Nf^2 \tr^2
-\frac{9}{32} \cf^2 \Nf \tr              \right.\right.\nonumber\\&{}&\left.\left.
+3 \cf \Nf^2 \tr^2   \zeta_3
-\frac{209}{48} \cf \Nf^2 \tr^2
+\frac{25 \Nf^3 \tr^3}{81}\right)            
+l_{\sss{\mu q}}^2 \left(
-\frac{1793   \ca^3}{216}          \right.\right.\nonumber\\&{}&\left.\left.
+\frac{273}{32} \ca^2 \Nf \tr
+\frac{55}{32} \ca \cf \Nf \tr   
-\frac{181}{72}   \ca \Nf^2 \tr^2            
-\frac{5}{8} \cf \Nf^2 \tr^2
+\frac{5 \Nf^3 \tr^3}{27}\right)                 \right.\nonumber\\&{}&\left.
+l_{\sss{\mu q}}^3   \left(
-\frac{1331 \ca^3}{1728}
+\frac{121}{144} \ca^2 \Nf \tr
-\frac{11}{36} \ca \Nf^2   \tr^2
+\frac{\Nf^3 \tr^3}{27}\right)                 \right.\nonumber\\&{}&\left.
\f{1}{\eps}\left[
+\frac{1415 \ca^2 \Nf \tr}{864 }
-\frac{2857 \ca^3}{1728 }
+\frac{205 \ca   \cf \Nf \tr}{288 }
-\frac{79 \ca \Nf^2 \tr^2}{432 }
-\frac{\cf^2 \Nf \tr}{16 }
-\frac{11 \cf \Nf^2 \tr^2}{72 }   \right]              \right.\nonumber\\&{}&\left.
\f{1}{\eps^2}\left[
-\frac{89 \ca^2 \Nf \tr}{144   }
+\frac{187 \ca^3}{288 }
-\frac{11 \ca \cf \Nf \tr}{48 }
+\frac{5 \ca \Nf^2   \tr^2}{36 }
+\frac{\cf \Nf^2   \tr^2}{12 } \right] 
\right\}\nonumber
\eea
The tree-level, one-loop and two-loop terms in \eqref{C1GG_3l_full} have been computed in
\cite{Novikov_scalargluonium}, \cite{Bagan:1989vm,Harnett:2004pg} and \cite{Zoller:2012qv} correspondingly.

As already observed at two-loop level \cite{Zoller:2012qv} there are divergent contact terms in $C_{1}^{\ssst{GG}}$
starting from $\mathcal{O}(\als^2)$. It is intersting to observe that these divergent terms can be expressed through
the $\beta$-function coefficients from \eqref{be:def}:\footnote{In the first version of this paper the sign in
the last term of \eqref{C1GGthroughbeta} was wrong. I thank \mbox{V.~Prochazka} and R.~Zwicky for pointing this out. 
As a consequence the power of $Z_G$ had to be changed from $-1$ to $1$ in \eqref{C1GGthroughbeta2}.}
\be
   C_{1}^{\ssst{GG}} = \f{1}{\eps}\left[-\as^2\,\beta_1-\as^3\,2\beta_2\right]
   + \f{1}{\eps^2}\left[+\as^3\,\beta_0\beta_1\right]+\text{finite} \label{C1GGthroughbeta}
   \ee
This feature points to the possibility
that the contact terms and hence the additive part of the renormalization of the Wilson coefficient $C_{1}^{\ssst{GG}}$
could be expressed completely through the $\beta$-function. 
An explanation for this curious behaviour and its meaning for the $O_1$-correlator remains to
be found. However, we can try to find a minimal closed formula for
the representation \eqref{C1GGthroughbeta} of the divergent part of $C_{1}^{\ssst{GG}}$.
A reasonable possibility reproducing \eqref{C1GGthroughbeta} to the given order in $\as$ is
\be \begin{split} C_{1}^{\ssst{GG}} 
&=\as^2 \lb 1-\f{\beta(\as)}{\eps}\rb^{-1} \f{\p}{\p \as}\left[
\f{\beta(\as)}{\eps\,\as}\right]+\mathcal{O}(\as^4)+\text{finite} \\
&= \as^2 Z_G \f{\p}{\p \as}\left[
\f{\beta(\as)}{\eps\,\as}\right]+\mathcal{O}(\as^4)+\text{finite} 
{}, \end{split} \label{C1GGthroughbeta2}
\ee
which contains a second order derivative of $Z_{\als}$ wrt $\als$.
It can be hoped that an explanation for this can be found along the lines of \cite{Spiridonov:1984br},
where the renormalization constant $Z_G$ in \eqref{ZGdef}
was obtained by taking first order derivatives of the generating functional of QCD wrt $\als$,
the gauge parameter and the external currents. We hope to return to this question in a future publication.

{\bf Note added 17th March 2016:} {\it The phenomenon of these contact terms has now been explained in \cite{Zoller:2016iam} where \eqref{C1GGthroughbeta2}
has been confirmed and even proven to be an exact identitity valid to all orders in $\als$.}

An unambiguous result can be obtained for the Adler function of $C_{1}^{\ssst{GG}}$, in which all contact terms,
finite and divergent, vanish:
\bea
 Q^2\f{d}{dQ^2}\, C_1^{\ssst{GG}}= &{}&
 \as \left\{\frac{11 \ca}{12}-\frac{\Nf \tr}{3}\right\}\nonumber\\
 &+&\as^2 \left\{
\frac{1151   \ca^2}{216}
-\frac{97 \ca \Nf \tr}{27}
-\cf \Nf \tr+\frac{10 \Nf^2   \tr^2}{27} \right.\nonumber\\&{}&\left.
+l_{\sss{\mu q}}   \left(
\frac{121 \ca^2}{72}
-\frac{11 \ca \Nf \tr}{9}
+\frac{2 \Nf^2 \tr^2}{9}
\right)
\right\}\nonumber\\
&+&\as^3 \left\{
-\frac{363 \ca^3 \zeta_3}{32}
+\frac{360325 \ca^3}{10368}
-\frac{55757 \ca^2 \Nf   \tr}{1728} \right.\nonumber\\&{}&\left.
+\frac{33}{4} \ca \cf \Nf   \tr \zeta_3       
-\frac{2527}{192} \ca \cf \Nf \tr
+\frac{3}{2} \ca \Nf^2 \tr^2   \zeta_3 \right.\nonumber\\&{}&\left.
+\frac{2057}{288} \ca \Nf^2 \tr^2
+\frac{9}{32} \cf^2 \Nf \tr          
-3 \cf \Nf^2 \tr^2   \zeta_3 \right.\nonumber\\&{}&\left.
+\frac{209}{48} \cf \Nf^2 \tr^2
-\frac{25 \Nf^3 \tr^3}{81}         
+l_{\sss{\mu q}} \left(
\frac{1793 \ca^3}{108}    \right.\right.\nonumber\\&{}&\left.\left.
-\frac{273}{16} \ca^2 \Nf \tr    
-\frac{55}{16} \ca   \cf \Nf \tr 
+\frac{181}{36} \ca \Nf^2 \tr^2\right.\right.\nonumber\\&{}&\left.\left.
+\frac{5}{4} \cf \Nf^2 \tr^2
-\frac{10   \Nf^3 \tr^3}{27}\right)  
+l_{\sss{\mu q}}^2 \left(
\frac{1331 \ca^3}{576} \right.\right.\nonumber\\&{}&\left.\left.
-\frac{121}{48} \ca^2 \Nf   \tr
+\frac{11}{12} \ca \Nf^2 \tr^2
-\frac{\Nf^3 \tr^3}{9}\right)
\right\}\label{C1GG_Adler}
\eea
In analogy to the construction above we can also find an RGI Wilson coefficient
\be C_1^{\ssst{GG,RGI}}:= \hat{\beta}( a_s) \, C_1^{GG}, \ee
which fulfills
\be C^{\ssst{GG,RGI}}_{1} O_1^{\ssst{RGI}}=C^{\ssst{GG}}_{1}[O_1]. \ee
For the derivative of the Wilson coefficient wrt $Q^2$ we find 
\bea
 Q^2\f{d}{dQ^2}\, C_1^{\ssst{GG,RGI}}= &{}&
    \as^2 \left\{\frac{11 \ca}{12}-\frac{\Nf \tr}{3}\right\}\nonumber\\
       &+&\as^3 \left\{
   \frac{163   \ca^2}{27}
   -\frac{433 \ca \Nf \tr}{108}
   -\frac{5 \cf \Nf \tr}{4}
   +\frac{10 \Nf^2   \tr^2}{27}\right.\nonumber\\&{}&\left.
   +l_{\sss{\mu q}}   \left(
   \frac{121 \ca^2}{72}
   -\frac{11 \ca \Nf \tr}{9}
   +\frac{2 \Nf^2 \tr^2}{9}\right)
   \right\}\label{C1GG_Adler_RGI}\\
   &+&\as^4\, \f{1}{(11 \ca-4 \Nf \tr)} \left\{
   -\frac{3993 \ca^4 \zeta_3}{32 }
   +\frac{565933 \ca^4}{1296}
   +\frac{363 \ca^3 \Nf \tr \zeta_3}{8 }\right.\nonumber\\&{}&\left.
   -\frac{730223 \ca^3   \Nf \tr}{1296 }
   +\frac{363 \ca^2 \cf \Nf \tr \zeta_3}{4 }
   -\frac{16625 \ca^2 \cf \Nf \tr}{96 }\right.\nonumber\\&{}&\left.
   +\frac{33   \ca^2 \Nf^2 \tr^2 \zeta_3}{2 }
   +\frac{100667 \ca^2 \Nf^2 \tr^2}{432 }
      +\frac{7 \cf^2 \Nf^2 \tr^2}{4 }
   +\frac{55 \ca \cf^2 \Nf \tr}{16}\right.\nonumber\\&{}&\left.
   +12   \cf \Nf^3 \tr^3 \zeta_3
   -\frac{113 \cf \Nf^3 \tr^3}{6 }
   -66 \ca \cf \Nf^2 \tr^2 \zeta_3
   +\frac{1423 \ca \cf \Nf^2 \tr^2}{12 }\right.\nonumber\\&{}&\left.
   +\frac{100 \Nf^4   \tr^4}{81 }
   -6 \ca \Nf^3 \tr^3 \zeta_3 
   -\frac{11075 \ca \Nf^3 \tr^3}{324 }\right.\nonumber\\&{}&\left.
   +l_{\sss{\mu q}} \left(
   \frac{85063 \ca^4}{432 }
   -\frac{117887   \ca^3 \Nf \tr}{432 }
   -\frac{2057 \ca^2 \cf \Nf \tr}{48 }
   +\frac{1184 \ca^2 \Nf^2 \tr^2}{9 }\right.\right.\nonumber\\&{}&\left.\left.
   -\frac{17 \cf   \Nf^3 \tr^3}{3 }
   +\frac{187 \ca \cf \Nf^2 \tr^2}{6 }
   +\frac{40 \Nf^4 \tr^4}{27 }
   -\frac{683 \ca \Nf^3   \tr^3}{27 }\right)\right.\nonumber\\&{}&\left.
   +l_{\sss{\mu q}}^2 \left(
   \frac{14641 \ca^4}{576 }
   -\frac{1331 \ca^3 \Nf \tr}{36 }
   +\frac{121 \ca^2 \Nf^2   \tr^2}{6 }
   +\frac{4 \Nf^4 \tr^4}{9 }
   -\frac{44   \ca \Nf^3 \tr^3}{9}
   \right)
   \right\}.\nonumber
\eea

\section{Numerics}

Finally, we consider two cases which are interesting for applications numerically,
that is gluodynamics ($n_f=0$) and QCD with only three light quarks ($n_f=3$). 
For this we choose the scale $\mu^2=Q^2$, i.e. we set $l_{\sss{\mu q}} = 0$.
For the correlator \eqref{TThat} we find

\bea
C^{(S)}_{1}(\mu^2=Q^2, \Nf=0)  &=&  \frac{22}{9}\as \left( 1+0.943182 \,\as - 7.06061 \,\as^2\right)
{},\\
C^{(S)}_{1}(\mu^2=Q^2, \Nf=3)  &=&  2\as \left( 1 + 0.652778 \as - 5.18519 \as^2\right)
{},\\
C^{(T)}_{1}(\mu^2=Q^2, \Nf=0)  &=&  -\frac{5}{6}\as  \left( 1  +2.075 \as + 14.3904 \as^2\right)
{},\\
C^{(T)}_{1}(\mu^2=Q^2, \Nf=3)  &=&  -\frac{15}{16}\as  \left( 1 +0.444444 \as + 6.64113 \as^2 \right)
{}.
\eea
and for \eqref{O1O1:def} we get
\bea
Q^2\f{d}{dQ^2}\,C_1^{\ssst{GG}}(\mu^2=Q^2, \Nf=0) &=&  \f{11}{4} \as^2
\left(1    +17.4394 \as + 207.338 \as^2
\right )
{}, \\
Q^2\f{d}{dQ^2}\,C_1^{\ssst{GG}}(\mu^2=Q^2, \Nf=3)  &=&  \f{9}{4} \as^2
\left(1   +13.6111 \as + 78.8642 \as^2
\right )
{}. 
\eea
For the RGI coefficients the numerical evaluation yields
\bea
C^{(S)}_{1,\ssst{RGI}}(\mu^2=Q^2, \Nf=0)  &=&  \frac{22}{9} \left( 1- 1.375 \, \as -11.9896\, \as^2\right)
{},\\
C^{(S)}_{1,\ssst{RGI}}(\mu^2=Q^2, \Nf=3)  &=&  2 \left( 1 - 1.125  \, \as -7.65625\, \as^2\right)
{},\\
C^{(T)}_{1,\ssst{RGI}}(\mu^2=Q^2, \Nf=0)  &=&  -\frac{5}{6} \left( 1 -  0.2431825 \, \as +6.83767\, \as^2\right)
{},\\
C^{(T)}_{1,\ssst{RGI}}(\mu^2=Q^2, \Nf=3)  &=&  -\frac{15}{16} \left( 1 - 1.33333 \, \as +4.54043\, \as^2\right)
{}
\eea
and
\bea
Q^2\f{d}{dQ^2}\,C_1^{\ssst{GG},\ssst{RGI}}(\mu^2=Q^2, \Nf=0) &=&  \f{11}{4} \as
\left(1   + 19.7576 \, \as +255.882\, \as^2
\right )
{}, \\
Q^2\f{d}{dQ^2}\,C_1^{\ssst{GG},\ssst{RGI}}(\mu^2=Q^2, \Nf=3)  &=&  \f{9}{4} \as
\left(1   + 15.3889 \, \as +107.533 \as^2
\right )
{}.
\eea

The numerical impact of the higher order corrections can be seen by evaluating the RGI
coefficients at $\mu=M_Z$, $\mu=3.5$ GeV and $\mu=2$ GeV, where
\be \als^{(\Nf=5)}(M_Z)\approx 0.118 \text{ , } \als^{(\Nf=3)}(3.5 \text{GeV})\approx 0.24
\text{ and } \als^{(\Nf=3)}(2 \text{GeV})\approx 0.30 \text{ \cite{Chetyrkin:2000yt}} \label{alsvalues1}\ee
for the cases $\Nf=5$ and $\Nf=3$ respectively. We find
\begin{align}
C^{(S)}_{1,\ssst{RGI}}(Q^2=\mu^2=M_Z^2,\Nf=5)&= \frac{46}{27}\;\left(
-\underbrace{0.00705235}_{\text{3 loop}} 
-\underbrace{0.0359955}_{\text{2 loop}}  
+\underbrace{1}_{\text{1 loop}}\right) ,\\
C^{(S)}_{1,\ssst{RGI}}(Q^2=\mu^2=(3.5 \text{ GeV})^2,\Nf=3)&= 2\;\left(
-\underbrace{0.0446826 }_{\text{3 loop}} 
-\underbrace{0.0859437}_{\text{2 loop}} 
+\underbrace{1}_{\text{1 loop}} \right),\\
C^{(S)}_{1,\ssst{RGI}}(Q^2=\mu^2=(2 \text{ GeV})^2,\Nf=3)&= 2\;\left(
-\underbrace{0.0698166}_{\text{3 loop}} 
-\underbrace{0.10743}_{\text{2 loop}} 
+\underbrace{1}_{\text{1 loop}} \right)
\end{align}
and
\begin{align}
C^{(T)}_{1,\ssst{RGI}}(Q^2=\mu^2=M_Z^2,\Nf=5)&= -\frac{145}{144}\;\left(
\underbrace{0.00930401}_{\text{3 loop}} 
-\underbrace{0.0640238}_{\text{2 loop}}  
+\underbrace{1}_{\text{1 loop}} \right),\\
C^{(T)}_{1,\ssst{RGI}}(Q^2=\mu^2=(3.5 \text{ GeV})^2,\Nf=3)&= -\frac{15}{16}\;\left(
\underbrace{0.0264984}_{\text{3 loop}} 
-\underbrace{0.101859}_{\text{2 loop}} 
+\underbrace{1}_{\text{1 loop}} \right),\\
C^{(T)}_{1,\ssst{RGI}}(Q^2=\mu^2=(2 \text{ GeV})^2,\Nf=3)&= -\frac{15}{16}\;\left(
\underbrace{0.0414038}_{\text{3 loop}} 
-\underbrace{0.127324}_{\text{2 loop}} 
+\underbrace{1}_{\text{1 loop}}\right) . 
\end{align}
for the correlator \eqref{TThat}. This shows that for the energy-momentum tensor
the Wilson coefficient $C_1^{(T)}$ is well convergent, even at \mbox{$\mu=2$ GeV}.
The three-loop approximation for $C_1^{(S)}$ at low scales is less good, but still
acceptable. At \mbox{$\mu=3.5$ GeV} the three-loop correction is $50\%$ of the two-loop
correction but both together are only a $12\%$ correction to the one-loop result.

For the correlator \eqref{O1O1:def} we find with
\be \als^{(\Nf=3)}(5 \text{GeV})\approx 0.213 \text{ \cite{Chetyrkin:2000yt}} \ee
in addition to \eqref{alsvalues1}:
\bea
&Q^2&\f{d}{dQ^2}\,C_1^{\ssst{GG},\ssst{RGI}}(Q^2=\mu^2=M_Z^2,\Nf=5)\nonumber\\ &=& 
\frac{23}{12}\as^2(\mu=M_Z)\;\left(
\underbrace{0.0074766}_{\text{3 loop}} 
+\underbrace{0.439205}_{\text{2 loop}}  
+\underbrace{1}_{\text{1 loop}} \right),\\
&Q^2&\f{d}{dQ^2}\,C_1^{\ssst{GG},\ssst{RGI}}(Q^2=\mu^2=(5 \text{ GeV})^2,\Nf=3)\nonumber\\ &=& 
\frac{9}{4}\as^2(\mu=5 \text{ GeV})\;\left(
\underbrace{0.494311}_{\text{3 loop}} 
+\underbrace{1.04337}_{\text{2 loop}} 
+\underbrace{1}_{\text{1 loop}} \right),\\
&Q^2&\f{d}{dQ^2}\,C_1^{\ssst{GG},\ssst{RGI}}(Q^2=\mu^2=(2 \text{ GeV})^2,\Nf=3)\nonumber\\ &=& 
\frac{9}{4}\as^2(\mu=2 \text{ GeV})\;\left(
\underbrace{0.980582}_{\text{3 loop}} 
+\underbrace{1.46953}_{\text{2 loop}} 
+\underbrace{1}_{\text{1 loop}} \right). 
\eea
Here the convergence at low scales is not so good as the two-loop correction becomes larger than the one-loop
correction at \mbox{$\mu=5$ GeV} and the three-loop correction shifts the result by another $50\%$
of the one-loop results. This suggests that higher order corrections should always be taken into account 
when this coefficient is used e.g. in sum rules
and special care has to be taken with regard to the convergence of the perturbation series at the scale
where perturbative and non-perturbative physics are separated in the OPE.
With this in mind, extending $C_1^{\ssst{GG}}$ to even higher orders in the future
could therefore be an interesting task.

\section{Conclusions}

We have presented the missing three-loop corrections to the OPE of the correlator of two
scalar gluonic operators $[O_1]=-\f{Z_G}{4}G^{\ssst{B} a\,\mu\nu}G^{\ssst{B} a}_{\mu\nu}$
and of the correlator of two energy-momentum tensors $T^{\mu\nu}$ in massless QCD at zero temperature.

These are the three-loop contributions to the coefficient $C_1$ in front of the local operator $[O_1]$.
We have also constructed renormalization group invariant versions of these coefficients and confirmed
the predictions made in \cite{Zoller:2012qv} for the logarithmic part of these coefficients.

In the coefficient $C_1^{\ssst{GG}}$ for the $O_1$-correlator we observe the curious feature that
divergent contact terms appear which are expressible through the QCD $\beta$-function.
These constact terms as well as contact terms in the $T^{\mu\nu}$-correlator proportional to
the tensor structures $t_4^{\mu\nu;\rho\sigma}(q)$ and $t_5^{\mu\nu;\rho\sigma}(q)$ from
\eqref{t1_t5} have to be subtracted. If we consider only derivatives wrt $Q^2$ of ambiguous
Wilson coefficients these terms vanish automatically.

All results can be found in a machine-readable form at\\
\texttt{\bf http://www-ttp.particle.uni-karlsruhe.de/Progdata/ttp14/ttp14-023/}

\section*{Acknowledgments}
I thank K.~G.~Chetyrkin for many useful discussions and his collaboration on 
the previous project \cite{Zoller:2012qv}.
I am also grateful to J.~H.~K\"uhn for his support and useful comments. 
This work has been supported by the Deutsche Forschungsgemeinschaft in the
Sonderforschungsbereich/Transregio SFB/TR-9 ``Computational Particle
Physics'', the Graduiertenkolleg ``Elementarteilchenphysik
bei h\"ochsten Energien und h\"ochster Pr\"azission'' 
and the ``Karlsruhe School of Elementary Particle and Astroparticle Physics: Science and Technology (KSETA)''.

\bibliographystyle{JHEP}

\bibliography{Literatur_v2}

\end{document}